\documentclass[superscriptaddress,aps,prl,amssymb,amsfonts,letterpaper,twocolumn]{revtex4}
\usepackage{amsmath} \usepackage{amsthm} \usepackage{color}
\usepackage{graphics,graphicx}
\usepackage{hyperref}

\def\tr{{\mathsf{tr}}}

\def\({\left(}
\def\){\right)}

\def\cW{{\mathcal W}}

\newcommand{\ket}[1]{| #1 \rangle}
\newcommand{\bra}[1]{\langle #1 |}

\newcommand{\hil}{\mathcal{H}}
\newcommand{\bop}{\mathcal{B}}

\def\01{\{0,1\}}

\begin{document}

\title{Local quantum measurement and no-signaling imply quantum correlations}
\author{H.~Barnum}
\affiliation{Perimeter Institute for Theoretical Physics, 31 Caroline St. N, 
Waterloo, ON, N2L 2Y5 Canada}

\author{S.~Beigi}
\affiliation{Institute for Quantum Information,
  California Institute of Technology, Pasadena, CA 91125, USA}

\author{S.~Boixo}
\affiliation{Institute for Quantum Information,
  California Institute of Technology, Pasadena, CA 91125, USA}
\email{boixo@caltech.edu}

\author{M.~B.~Elliott}
\noaffiliation

\author{S.~Wehner}
\affiliation{Institute for Quantum Information,
  California Institute of Technology, Pasadena, CA 91125, USA}

\date{\today}

\begin{abstract}

We show that, assuming that quantum mechanics holds locally, the
finite speed of information is the principle that limits all possible
correlations between distant parties to be quantum mechanical as
well. Local quantum mechanics means that a Hilbert space is assigned
to each party, and then all local POVM measurements are (in principle)
available; however, the joint system is not necessarily described by a
Hilbert space. In particular, we do not assume the tensor product
formalism between the joint systems. Our result shows that if any
experiment would give non-local correlations beyond quantum mechanics,
quantum theory would be invalidated even locally.
\end{abstract}

\maketitle

Quantum correlations between space-like separated systems are, in the words of Schr\"odinger, ``\emph{the} characteristic trait of quantum mechanics, the one that enforces its entire departure from classical lines of thought''\cite{schrdinger_discussion_1935}.
Indeed, the increasing experimental support~\cite{experiments}
for correlations violating Bell inequalities~\cite{bell_einstein_1964}
is at odds with local realism.
Quantum correlations have been investigated with
increasing success~\cite{quantum-bg}, but what is the principle that limits them~\cite{popescu_quantum_1994}?

Consider two experimenters, Alice and Bob, at two distant
locations. They share a preparation of a bipartite physical system, on
which they locally perform one of several measurements. This shared
preparation may thereby cause the distribution over the possible two
outcomes to be correlated.  In nature, such non-local correlations
cannot be arbitrary.  For example, it is a consequence of relativity
that information cannot propagate faster than light.  The existence 
of a finite upper bound on the speed of information is known as
the principle of \emph{no-signaling}.  This principle implies that if
the events corresponding to Alice's and Bob's measurements are separated
by space-like intervals, then Alice cannot send information to Bob by
just choosing a particular measurement setting. Equivalently, the
probability distribution over possible outcomes on Bob's side cannot
depend on Alice's choice of measurement setting, and vice
versa. Quantum mechanics, like all modern physical theories, obeys the
principle of no-signaling.

But is no-signaling the only limitation for correlations observed in
nature?  Bell~\cite{bell_einstein_1964} initiated the study of these
limitations based on inequalities, such as the CHSH
expression~\cite{clauser_proposed_1969}.  It is convenient to describe
this inequality in terms of a game played by Alice and Bob.  Suppose
we choose two bits $x,y \in \01$ uniformly and independently at
random, and hand them to Alice and Bob respectively. We say that the
players win, if they are able to return answers $a, b \in \01$
respectively, such that $x \cdot y = a + b \mod 2$. Alice and Bob can
agree on any strategy beforehand, that is, they can choose to share
any preparation possible in a physical theory, and choose any
measurements in that theory, but there is no further exchange of
information during the game.  The probability that the players win is
\begin{align}
 \frac{1}{4} \sum_{x,y \in \01} \sum_{\substack{a,b \in \01\\ x \cdot y = a + b \mod 2}} p(a,b|M_A^x, M_B^y)
\end{align}
where $p(a,b|M_A^x, M_B^y)$ denotes the probability that Alice and Bob obtain measurement outcomes $a$ and $b$
when performing the measurements $M_A^x$ and $M_B^y$ respectively (any pre- or post-processing can be taken as part of the measurement operation).
Classically, i.e. in any local realistic theory, this probability is bounded by~\cite{clauser_proposed_1969}
\begin{align}\label{eq:classic}
  p_{\rm classical} \le 3/4\;.
\end{align}
Such an upper bound is called a Bell inequality.

Crucially, Alice and Bob can violate this inequality using quantum mechanics
~\cite{bell_einstein_1964}.
The corresponding bound is~\cite{tsirelson_quantum_1980}
\begin{align}\label{eq:tsirel}
p_{\rm quantum} &\leq
\frac{1}{2} + \frac{1}{2 \sqrt{2}} \ ,
\end{align}
and there exists a shared quantum state and measurements that achieve it~\cite{clauser_proposed_1969}.
Further, there is now compelling experimental evidence that nature violates Bell inequalities and does not admit a local realistic description~\cite{experiments}.
Yet, there exist stronger no-signaling correlations (outside quantum mechanics) which achieve success
probability $p_{\rm no signal} = 1$~\cite{popescu_quantum_1994}. 
So why, then, isn't nature more non-local~\cite{popescu_quantum_2006}? 

Studying limitations on non-local correlations thus forms an essential element of understanding nature. On one hand, it provides a systematic method to both theoretically and experimentally compare candidate physical theories~\cite{leggett}. On the other hand, it crucially affects our understanding of information in different settings such as cryptography and communication complexity~\cite{settings,wim:nonlocal,gs:relaxedUR, pawlowski_new_2009}.
For example, if nature would admit $p_{\rm no signal} = 1$, any two-party communication problem could be solved using only a single bit of communication, independent of its size~\cite{wim:nonlocal}.
Also, for the special case of the CHSH inequality, it is known that the bound~\eqref{eq:tsirel}
is a consequence of information theoretic constraints such as uncertainty relations~\cite{gs:relaxedUR} or
the recently proposed principle of information causality~\cite{pawlowski_new_2009}.
However, characterizing general correlations remains a difficult challenge~\cite{challenge}, and it is interesting
to consider what other constraints may impose limits on quantum correlations.

%\section{Result}

\begin{figure*}[ht!]
\begin{center}
\includegraphics[scale=0.7]{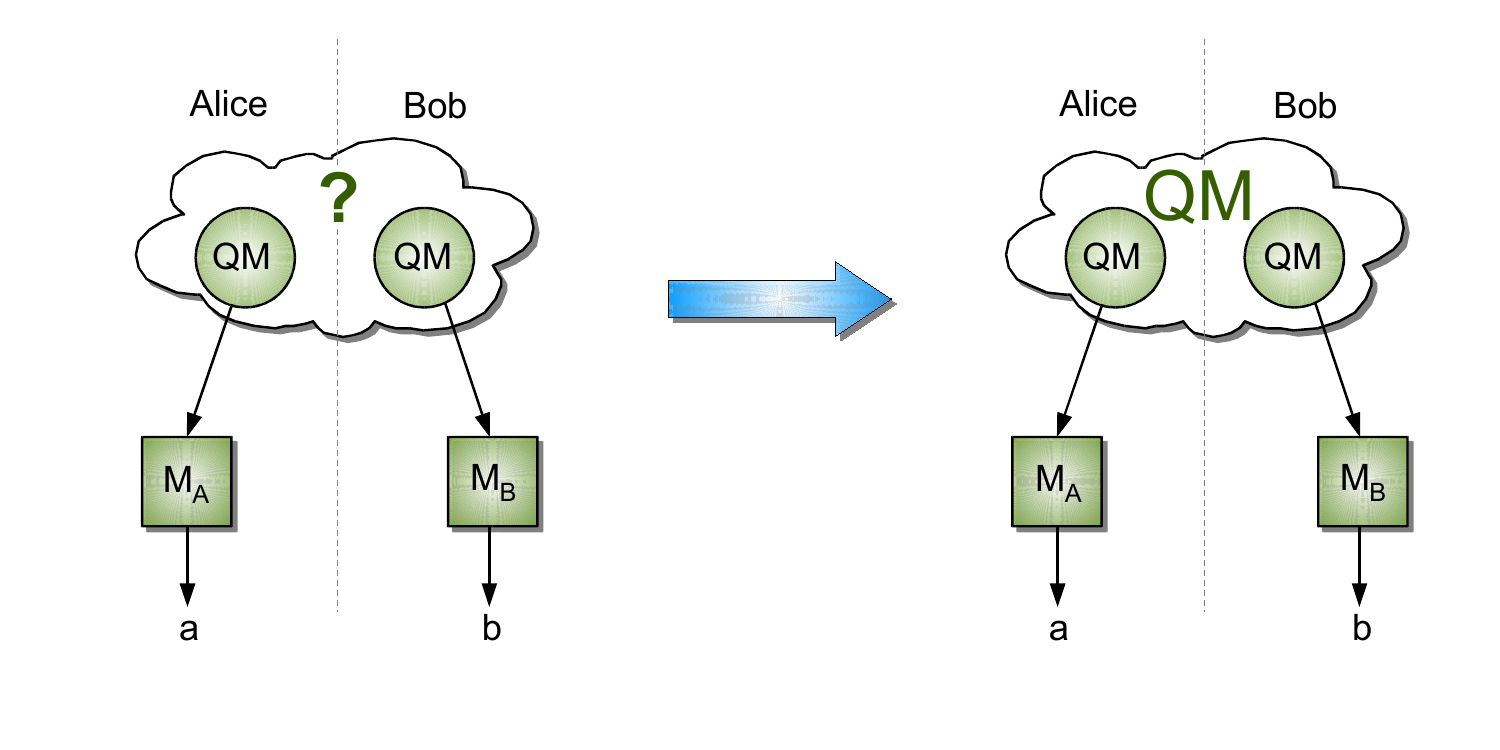}
\caption{If the principle of no-signaling is obeyed and Alice and Bob are locally quantum, their non-local correlations
can be obtained in quantum mechanics. Alice and Bob are \emph{locally quantum} if their local systems can be described by a Hilbert space and they can choose to measure \emph{any} local POVM
$M_A = \{Q_a\}_a$ and $M_B = \{R_b\}_b$. A shared preparation between Alice and Bob corresponds to a function $\omega$ on the pair of POVM elements such that $p(a,b|M_A,M_B)=\omega(Q_a,R_b)$.
This, with no-signaling, implies that
the marginal distributions are given by Born's rule, $p(a|M_A) = \tr(Q_a \rho_A)$ and $p(b|M_B) = \tr(R_b \rho_B)$,
where $\rho_A$ and $\rho_B$ are quantum states (but the state of their joint system may not be quantum). We show that, in this setting, for any preparation $\omega$ there exist a joint quantum state $\sigma_{AB}$ and a relabeling of POVM measurements
$\{\tilde{M}_A\}$ and $\{\tilde{M}_B\}$, such that
$\omega(Q_a,R_b)=p(a,b|M_A,M_B) = \tr((\tilde Q_a \otimes \tilde R_b)\sigma_{AB})$ where $\tilde{M}_A = \{\tilde Q_a\}_a$ and $\tilde{M}_B = \{\tilde R_b\}_b$.
}
\label{fig:result}
\end{center}
\end{figure*}

{\bf Result.} We forge a fundamental link between local quantum theory and non-local quantum correlations.
In particular, we show that if Alice and Bob are \emph{locally quantum},
then relativity theory implies that their non-local correlations admit a quantum description. 
The assumption of being locally quantum may thus provide another ``reason'' why the correlations we observe in nature
are restricted by more than the principle of no-signaling itself. Figure~\ref{fig:result} states our result.

Let us explain more formally what we mean by being \emph{locally
  quantum} (see also Figure~\ref{fig:result}).  We say that Alice is
locally quantum, if her physical system can be described by means of a
Hilbert space $\hil_A$ of some fixed finite dimension $d$, on which
she can perform \emph{any} local quantum measurement (POVM) $M_A =
\{Q_a\}_a$ given by bounded operators $Q_a \in \bop(\hil_A)$. The
probability $p(a|M_A)$ that she obtains an outcome $a$ for measurement
$M_A = \{Q_a\}_a$ is given by a function $ \bop(\hil_A) \rightarrow
[0,1]$ applied to the POVM elements.  Gleason's theorem for POVM
elements implies that the state of Alice's system is then described by
a state $\rho_A \in \bop(\hil_A)$~\cite{gleason}, and similarly for Bob,
where we use $\hil_B$ and $M_B = \{R_b\}_b$ to denote his Hilbert
space and measurements respectively.  Conceptually, this means that
quantum mechanics describes Alice and Bob's local physical systems.

However, we make no a priori assumption about the nature of the joint system held by Alice and Bob. In particular, we do not assume that it is 
described by a tensor product of their local Hilbert spaces, or that their joint system is quantum mechanical. 
This means that Alice and Bob can share any possible \emph{preparation} which assigns probabilities to \emph{local} POVM measurements. 
That is, their preparation is simply a function $\omega$ such that the probabilities of observing outcomes $a$ and $b$ for 
measurements $M_A = \{Q_a\}_a$ and $M_B = \{R_b\}_b$ are given by $p(a,b|M_A,M_B) = \omega(Q_a,R_b)$. In particular, the state of their joint system
may not be described by any density matrix.

Nevertheless, we are able to show that just from the assumptions that
Alice and Bob are locally quantum and that the no-signaling principle is
obeyed, it follows that there exist a Hilbert space $\hil_{AB} =
\hil_A \otimes \hil_B$, a state $\rho_{AB} \in \bop(\hil_{AB})$ and
measurements $\tilde{M}_A = \{\tilde{Q}_a\}_a$ and $\tilde{M}_B =
\{\tilde{R}_b\}_b$ for Alice and Bob, such that
\begin{align}
  p(a,b|M_A,M_B) = \omega(Q_a,R_b) = \tr((\tilde{Q}_a \otimes \tilde{R}_b)\rho_{AB}) 
\end{align}
That is, all correlations can be reproduced
quantum mechanically.

{\bf Implications.}
% In spirit, our result is analogous to the classical case, where if 
% Alice and Bob's local measurements are classical and their correlations are no-signaling, then
% their correlations admit a classical description.
% Here we do the same for local \emph{quantum} measurements and no-signaling. 
Our result solves an important piece of the puzzle of understanding
non-local correlations, and their relation to the rich \emph{local} phenomena we encounter in quantum theory such as Bohr's complementarity principle, Heisenberg uncertainty
and Kochen-Specker
non-contextuality.
In particular, it implies that if we obey local quantum statistics
we can never hope to surpass a Tsirelson-type bound on $p_{\rm quantum}$ like that of~\eqref{eq:tsirel}, ruling out the possibility
of such striking differences with respect to information processing as those pointed out in~\cite{wim:nonlocal}. 
Indeed, \emph{if} we were able to surpass such bounds,
then the \emph{local} systems of Alice and Bob could not be quantum.

Other recent works also attempt to explain the limitations of quantum
correlations.  For example, the principle of information
causality~\cite{pawlowski_new_2009} starts with the assumption that
nature demands that certain communication tasks should be hard to
solve. Together with the assumption of the no-signaling principle,
this allows one to obtain Tsirelson's bound for the special case of
the CHSH inequality. In our work, we also assume the no-signaling
principle, but combine it with a different assumption, namely, that
the world is locally quantum, that is, quantum mechanics correctly
describes the laws of nature of local physical systems. Making this
assumption we recover the quantum limit on \emph{all} possible
non-local correlations (not only the Tsirelson's bound for the CHSH
inequality).

% We would like to emphasize that our result does not rule out the \emph{existence} of shared preparations that go beyond
% quantum states. Our goal is to show that by weakening the assumption of quantum physics to local systems we do not obtain stronger non-local correlations. 
% Indeed, we will see that one cannot rule out the existence of preparations beyond density matrices only by studying non-local 
% correlations, and hence even principles
% like information causality can never be used to single out quantum mechanics as a physical theory. 

%\section{Proof}

{\bf Proof.} To prove our result, we now proceed in two steps. 
First, we explain a known characterization of all no-signaling probability assignments to local quantum measurements~\cite{foulis_empirical_1979,klay_tensor_1987,barnum_influence-free_2005}.
Second, we use this characterization to show that the resulting correlations can be obtained in quantum mechanics.

%\subsection{From local quantum measurements to POPT states}

\emph{From local quantum measurements to POPT states:}
Fix two finite dimensional Hilbert spaces on Alice and Bob's sides. A local quantum measurement (or POVM) consists of a pair of measurements $M_A$ and $M_B$ with outcome labels $\{a\}$ and $\{b\}$ respectively on Alice and Bob's Hilbert spaces. Such POVMs are described by complex Hermitian matrices $M_A=\{Q_a\}_a$, $M_B=\{R_b\}_b$, $Q_a, R_b\ge 0$, which sum to the identity, i.e., $\sum_a Q_a = \sum_b R_b =\openone$ (see Figure~\ref{fig:result}). A preparation shared between Alice and Bob assigns outcome probabilities $p(a,b|M_A,M_B)$ to any choice of measurements $M_A$ and $M_b$. More precisely, it corresponds to a function $\omega$ on the pair of POVM elements such that $p(a,b|M_A,M_B)=\omega(Q_a, R_b)$.

Kl\"ay, Randall, and Foulis~\cite{klay_tensor_1987} have shown (see Appendix) that assuming no-signaling, the shared preparations (or equivalently the functions $\omega$) are in one-to-one correspondence with matrices $W_{AB}$ such that $\tr(W_{AB})=1$ and~\footnote{In the same reference it is shown that classical communication in the performance of measurements rules out signaling in the direction opposite to the classical communication.}
\begin{align}
\label{eq:popt}
  p(a,b|M_A,M_B) = \tr \( \( Q_a \otimes R_b \) W_{AB} \) \ge 0\;.
\end{align}
The matrices $W_{AB}$ are called \emph{positive on pure tensors} (POPT) states. All quantum states are POPT states, but there are POPT states that do not correspond to quantum 
states~\footnote{Note that our result does not rule out the
  existence of POPT states, but shows that they do not generate
  stronger than quantum correlations. }.

Note that POPT states cannot be combined
arbitrarily~\cite{barnum_influence-free_2005}.  For example, not all
entangled measurements (measurements which are not a convex
combination of tensor products $Q_a \otimes R_b$) of POPT states are
well defined because they would result in negative ``probabilities''
for non-quantum POPTs.  Specifically, if Alice and Bob share a POPT,
and Charlie and Bob share another one, then if Alice and Charlie come
together, entangled measurements between their POPTs are not
necessarily defined.  This does not affect our result, since we are
only interested in the case where we consider parties (here Alice and
Charlie together) which are locally quantum.
% Note however that based on our result the correlations between Alice and Bob, and Charlie and Bob remain the same even when combining POPT states, because they always admit a quantum description.

%\subsection{From POPT states to quantum correlations}

\emph{From POPT states to quantum correlations:} 
We now show that there exist a quantum state $\sigma_{AB}$ and a map on POVM measurements
\begin{align}
  f : \{\,M_A=\{Q_a\}_a \}  \mapsto  \{\, \tilde M_A = \{\tilde Q_a\}_a \}
\end{align}
such that
\begin{align}\label{eq:result}
  p(a,b|M_A, M_B) = \tr \((\tilde Q_a \otimes R_b) \sigma_{AB}\)\;.
\end{align}
In order to do so, we associate to each POPT state $W_{AB}$ a map $\cW$ from matrices to matrices using the Choi-Jamio{\l}kowski isomorphism. 
Explicitly, $W_{AB}$ is obtained from $\cW$ by acting on Bob's side of the (projection on the) maximally entangled state $\ket \Phi$
\begin{align}
  W_{AB} = \openone \otimes \cW (\ket \Phi \bra \Phi)\;.
\end{align}

Because $W_{AB}$ is a POPT, the associated map $\cW$ is positive, i.e., it sends positive matrices to positive ones, but it may not be an admissible quantum operation. Nevertheless, if $\cW$ still maps POVMs to POVMs we can obtain the POPT correlations by moving the action of $\cW$ from the maximally entangled state to the measurement elements. In particular, if  $\cW$ is unital ($\cW(\openone) = \openone$), the map
\begin{align}\label{eq:opDef}
f:Q_a \mapsto \tilde Q_a = \cW(Q_a^T)^T\ ,
\end{align}
maps POVM measurements to POVM measurements.
We then show that~\eqref{eq:result} holds with $\sigma_{AB}=\ket \Phi \bra \Phi$.
Let $d$ be the local dimension of Alice and Bob. If $\cW$ is unital we have
\begin{align}
  \tr \left ( (Q_a \otimes R_b) W_{AB}\right)&=
\tr \left((Q_a \otimes R_b)\openone \otimes \cW (\ket \Phi \bra \Phi) \right)\nonumber \\
  &= \tr \left(\ket \Phi \bra \Phi (\openone \otimes \cW^*) (Q_a \otimes R_b)\right)\nonumber \\
  &= \tr \left(\ket \Phi \bra \Phi\, (Q_a \otimes \cW^*(R_b))\right)\nonumber \\
  &= \frac 1 d \tr \left(Q_a^T \cW^*(R_b)\right)\nonumber \\
  &= \frac 1 d \tr \left(\cW(Q_a^T) R_b\right)\nonumber\\
  &= \tr \left(  (\tilde Q_a \otimes R_b) \ket \Phi \bra \Phi  \right)\;,
\end{align}
where $\cW^*$ denotes the adjoint of $\cW$. This establishes~\eqref{eq:result} in the unital case.

In general, $\cW$ can be decomposed into a unital map and another map. This other map gives a quantum state $\sigma_{AB}$ by acting on $\ket \Phi$.
 Then $f$ is defined in terms of the unital map as before. We finish the proof by showing that $\sigma_{AB}$ is well-normalized and~\eqref{eq:result} is satisfied.
For a general positive map, let $M$ be the image of the identity, i.e., $\cW(\openone) = M$. The matrix $M$ is normalized, $\tr (M)/d = \tr (W_{AB}) =1$. We assume initially that $M$ is invertible, and define
\begin{align}
  \tilde \cW (\cdot) = M^{-1/2} \cW (\cdot) M^{-1/2}\;.
\end{align}
The map $\tilde \cW$ is unital. Further, the quantum state $\sigma_{AB} =\ket \psi\bra \psi$ given by
\begin{align}\label{eq:psi}
\ket \psi = ( M^{1/2})^T \otimes \openone \ket \Phi\;
\end{align}
is well-normalized, that is,
$\tr(\sigma_{AB}) =  \tr(M^T)/d = 1$.
Thus by defining $f$ as in~\eqref{eq:opDef} but in terms of $\tilde \cW$ we conclude
\begin{align}
   \tr \left((Q_a \otimes R_b) W_{AB}\right) & = \frac{1}{d}\tr \left( \cW(Q_a^T) R_b\right)\nonumber \\
   &= \frac{1}{d} \tr \(\( M^{1/2} \tilde \cW(Q_a^T)M^{1/2}\)  R_b\)\nonumber \\
   %&= \tr \( \ket \Phi \bra \Phi \( (M^{1/2})^T \tilde Q_a (M^{1/2})^T \) \otimes R_b\) \nonumber \\
   &= \tr \( (\tilde Q_a \otimes R_b) \sigma_{AB}\)\;.
\end{align}

If $M$ is not invertible, in order to define $\tilde \cW$, one can start with the map $(1-\epsilon) \cW (\cdot) + \epsilon \openone \tr(\cdot)$, and then take the limit $\epsilon \to 0$.

%\section{Conclusion}

{\bf Conclusion.} We have shown that being locally quantum is
sufficient to ensure that all non-local correlations between distant
parties can be reproduced quantum mechanically, if the principle of
no-signaling is obeyed.  This gives us a natural explanation of why
quantum correlations are weaker than is required by the no-signaling
principle alone, i.e., given that one can describe local physics
according to quantum measurements and states, then no-signaling
already implies quantum correlations.

It would be interesting to know whether our work can be used to derive
more efficient tests for non-local quantum correlations than those
proposed in~\cite{challenge}.  Finally, it is an intriguing question
whether one can find new limits on our ability to perform information
processing \emph{locally} based on the limits of non-local
correlations, which we now know to demand local quantum behavior.

%%%%%%%%%%%%%%%%%%%%%%%%%%%%%%%%%%%%%%%%

%{\bf Acknowledgments} 
\acknowledgments This work was supported by the National Science
Foundation under grant PHY-0803371 through the Institute for Quantum
Information at the California Institute of Technology, and by the US
Department of Energy through the LDRD program at Los Alamos National
Laboratory.  Research at Perimeter Institute is supported by the
Government of Canada through Industry Canada and by the Province of
Ontario through the Ministry of Research and Innovation.

\appendix
\section{Appendix}

We include a derivation of the POPT states for completeness. We follow the more general version in~\cite{barnum_influence-free_2005}. The outline is the following: using no-signalling, we apply Gleason's theorem on both sides, Alice and Bob. This implies that the no-signaling POPT state is bilinear on Alice and Bob measurements, which gives its form.

We denote the local POVMs by
$M_A = \{Q_a\}_a$ and $M_B = \{R_b\}_b$. The joint probability distribution is given by a function $\omega$ acting on POVM elements
\begin{align}
  p(a,b|M_A,M_B)=\omega(Q_a,R_b)\;. 
\end{align}
Notice that for any pair of POVMs
\begin{align}
  \sum_{a,b} \omega(Q_a,R_b) =1\;,
\end{align}
but $\omega$ is not assumed to be bilinear at this point. No-signalling implies that for all $M_B$
\begin{align}
  \sum_b \omega(Q_a,R_b)  &= \sum_b p(a,b|M_A,M_B)\\ \nonumber &=  p(a|M_A,M_B) =  p(a|M_A) = \omega(Q_a)\;.
\end{align}
That is, the marginal distribution is well defined. 

For any POVM element $Q_a$ on Alice's side we can define a corresponding function $\omega_a$ which acts on Bob's POVM elements. The function $\omega_a$ is defined by its action on any POVM element $R_b$ with the equation
\begin{align}
 \omega_a(R_b) = \omega(Q_a,R_b)\;. 
\end{align}
Notice that, for every POVM $M_B$ on Bob's side, no-signalling from Bob to Alice implies that
\begin{align}
   \sum_b \omega_a(R_b) = \sum_b \omega(Q_a,R_b) = \omega(Q_a)\;. 
\end{align}
Because $\omega_a$ adds to the constant value $\omega(Q_a)$ when it is summed over any POVM, we can use Gleason's theorem~\cite{gleason_measuresclosed_1957,busch_quantum_2003,caves_gleason-type_2004} to identify $\omega_a$ with an \emph{unnormalized} quantum state $\tilde \sigma_a$ on Bob's side. 
Specifically, for any POVM element $R_b$, we have
\begin{align}
  \omega_a(R_b) = \omega(Q_a,R_b) = \tr(\tilde \sigma_a R_b)\;.
\end{align}
The previous equation allows us to define, for any given POPT $\omega$, a map $\hat \omega$ from POVM elements $Q_a$ on Alice's side to unnormalized quantum states on Bob's side
\begin{align}
  \hat \omega(Q_a) = \tilde \sigma_a\;. 
\end{align}

Now choose an informationally complete POVM $M_B=\{R_b\}$ on Bob's side. Then $\hat \omega$ is given by the functions $\omega^b$ defined by
\begin{align}
  \omega^b(Q_a) = \omega(Q_a,R_b)  = \tr(\tilde \sigma_a R_b)\;. 
\end{align}
We use no-signalling from  Alice to Bob to apply Gleason's theorem to each function $\omega^b$ from the informationally complete POVM, as we did before with no-signalling in the other direction. The action of $\omega^b$ is then given by an unnormalized quantum state, which implies that it is linear. This proves that $\hat \omega$ is linear.

Once we have established the linearity of $\hat \omega$ we can identify it with the operator $\cW$ introduced in the text according to
\begin{align}
  \hat \omega(Q_a) = \frac 1 d \cW(Q_a^T)\;. 
\end{align}
Finally, we can write
\begin{align}
  \omega(Q_a,R_b) &= \tr(\hat \omega(Q_a)R_b) = \frac 1 d \tr(\cW(Q_a^T) R_b) \\ \nonumber &= \tr((Q_a \otimes R_b) W_{AB} )\;. 
\end{align}

\end{document}